\documentclass{aa}
\usepackage{graphicx}
\usepackage[varg]{txfonts}
\bibpunct{(}{)}{;}{a}{}{,} 
\usepackage[colorlinks=true,linkcolor=blue,urlcolor=blue,citecolor=blue]{hyperref} 
\usepackage{siunitx}
\usepackage[normalem]{ulem} 
\usepackage{lipsum}
\usepackage{ulem}

\usepackage                             {booktabs}
\usepackage								{bm}
\usepackage     [labelfont=bf]          {caption}
\usepackage                             {subcaption}
\usepackage     [rightcaption]          {sidecap}
\usepackage{xcolor}
\DeclareSIUnit\gram{g}
\DeclareSIUnit\centimetre{cm}
\DeclareSIUnit\erg{erg}
\DeclareSIUnit\year{yr}
\DeclareSIUnit\ev{eV}

\usepackage{xspace}
\newcommand{\eos}{EoS\@\xspace}
\newcommand{\eg}{e.g.\@\xspace}
\newcommand{\cf}{c.f.\@\xspace}
\newcommand{\ie}{i.e.\@\xspace}
\newcommand{\mesa}{\mbox{\textsc{MESA}}\xspace}
\newcommand{\arepo}{\mbox{\textsc{arepo}}\xspace}

\newcommand{\upsub}[1]{\sb{\mathrm{#1}}}
\begingroup\lccode`~=`_\lowercase{\endgroup\let~\upsub}
\AtBeginDocument{%
  \catcode`_=12
  \mathcode`_="8000
}

\begin{document}

\title{Common-envelope evolution with an asymptotic giant branch star}

\author{Christian~Sand\inst{1}\fnmsep\thanks{\email{christian.sand@h-its.org}} \and
       Sebastian~T.~Ohlmann\inst{2} \and
       Fabian~R.~N.~Schneider\inst{3,1} \and 
       R\"{u}diger~Pakmor\inst{4} \and 
       Friedrich~K.~R{\"o}pke\inst{1,5}
       }

\titlerunning{Common-envelope evolution with an AGB star}
\authorrunning{Christian~Sand et al.}

\institute{%
    Heidelberger Institut f\"{u}r Theoretische Studien,
    Schloss-Wolfsbrunnenweg 35, 
    69118 Heidelberg, Germany
  \and
    Max Planck Computing and Data Facility,
    Gießenbachstr. 2,
    85748 Garching, Germany
  \and
    Zentrum f\"ur Astronomie der Universit\"at Heidelberg,
    Astronomisches Rechen-Institut, 
    M\"{o}nchhofstr. 12-14, 
    69120 Heidelberg, Germany
  \and
    Max-Planck-Institut für Astrophysik,
    Karl-Schwarzschild-Str. 1,
    85748 Garching, Germany
  \and
    Zentrum f\"ur Astronomie der Universit\"at Heidelberg,
    Institut f\"ur Theoretische Astrophysik, 
    Philosophenweg 12,
    69120 Heidelberg, Germany
}

\date{Received ; accepted }

\abstract {Common-envelope phases are decisive for the evolution of
  many binary systems. Cases with
  asymptotic giant branch (AGB) primary stars are of particular interest because they are
  thought to be progenitors of various astrophysical transients.  In
  three-dimensional hydrodynamic simulations with the moving-mesh code
  \textsc{arepo}, we study the common-envelope evolution of a
  $1.0\,M_{\odot}$ early-AGB star with companions of different
  masses. Although the stellar envelope of an AGB star is less
  tightly bound than that of a red giant, we find that the release of
  orbital energy of the core binary is insufficient to eject more than
  about twenty percent of the envelope mass. Ionization energy that is
  released in the expanding envelope, however, can lead to complete
  envelope ejection. Because recombination proceeds largely at high
  optical depths in our simulations, it is likely that this effect
  indeed plays a significant role in the considered systems. The
  efficiency of mass loss and the final orbital separation of the core
  binary system depend on the mass ratio between the companion and the
  primary star. Our results suggest a linear relation between the
  ratio of final to initial orbital separation and this parameter.}


\maketitle

\section{Introduction}
\label{sec:introduction}

Common envelope (CE) phases, which were first proposed by \citet{paczynski1976a},
pose great challenges to binary stellar evolution models. The physical
mechanism of these short episodes, where two stellar cores orbit each
other inside a giant star's envelope, is still not well understood. 
Due to tidal drag, the core binary system transfers orbital
energy to the envelope material. This may lead to envelope ejection
leaving behind cataclysmic variables, close white-dwarf main-sequence
binaries, double white-dwarfs, or other close binary systems of
compact stellar cores. Phenomena arising from post-CE binaries include 
Type Ia supernovae 
\citep{iben1984a, ruiter2009a, toonen2012a}, and also classical novae
\citep{livio1990a}, X-ray binaries \citep{kalogera1998a, taam2000a,
  taam2010a}, white-dwarf mergers \citep{pakmor2010a, ruiter2013a}, and
gravitational wave sources \citep{belczynski2002a}. 
Moreover, the CE phase is thought to be responsible for
the shapes of some planetary nebulae \citep{dekool1990a,
  nordhaus2007a, hillwig2016a, bermudez2020a}. 
  
Parameterized descriptions that are typically employed
in classical stellar-evolution theory and population-synthesis
calculations introduce large uncertainties in the predicted rates of these fundamentally important events. 
This situation calls
for an improved modeling of common-envelope evolution (CEE), which
requires a better understanding of the underlying physics and, in particular, 
the mechanism of envelope ejection that eludes one-dimensional
modeling.

We note that CEE \citep[for a review, see][]{ivanova2013a} starts out with unstable
mass transfer and the loss of co-rotation in the progenitor binary
system, followed by a rapid inspiral of the secondary star toward
the core of the primary star. In this phase, which is sometimes referred to as
``plunge-in'', large parts of the orbital energy are thought to be
transferred to the envelope. Numerical simulations, however, fail to
achieve envelope ejection during the plunge-in when only orbital
energy release is considered \citep[e.g.,][]{livio1988a, terman1994a,
  rasio1996a, sandquist1998a, sandquist2000a, passy2012a, ricker2012a,
  kuruwita2016a, ohlmann2016a, ohlmann2016b, staff2016b, iaconi2017a}.  A gradual
transfer of orbital energy over longer time scales in a subsequent
``self-regulated spiral-in'' has been proposed to eventually expel all
envelope material \citep{meyer1979a,podsiadlowski2001a}.  This phase,
however, is difficult to model and to date its efficiency, and even its very existence , remains
uncertain. 

The CEE is widely accepted as a mechanism for the formation of observed
close binary systems \citep{izzard2012a}, but simulations indicate 
that envelope ejection is probably not powered by the release of orbital 
energy alone. It seems likely that other energy sources are tapped or
other mechanisms of energy transfer to the envelope gas come into
play. 
Ionization energy has been identified as an important contribution to the overall energy budget of stellar envelopes \citep[e.g.,][]{biermann1938a, paczynski1968a}.
\cite{han1994a} discussed the ionization energy of H and He as part of the internal energy of the envelope of asymptotic giant branch (AGB) stars, which was soon after applied to the study of CEE \citep{han1995a}.
In the CE phase, the stellar envelope expands and recombination processes progressively liberate the ionization energy. This has been proposed as a mechanism leading to
successful envelope ejection \citep{nandez2015a, nandez2016a,
  prust2019a, reichardt2020a}. \citet{sabach2017a}. \citet{grichener2018a} and
\citet{soker2018a}, however, question its relevance because instead
of being converted locally into kinetic energy of the gas, the
recombination energy might be transported away by convection \citep{wilson2019a,wilson2020a} or
radiation \citep[see][]{ivanova2018a}. Other mechanisms such as dust formation
\citep{glanz2018a,iaconi2020a,reichardt2020a}, accretion onto the in-spiraling star
\citep{chamandy2018a}, and the formation of jets \citep{shiber2019a}
have been proposed to aid envelope ejection.

Three-dimensional hydrodynamic simulations strive to elucidate CEE,
but they pose a severe multi-scale, multi-physics problem. In
particular, the wide range of spatial scales ranging from the stellar
core to the giant star's envelope renders it difficult to achieve 
sufficient numerical resolution. For this reason, past simulations
focused on scenarios involving red-giant (RG) primary stars, where the
scale challenges are less severe than in the case of AGB primaries. 
CEE involving AGB stars is
particularly interesting, because the resulting close binary star
contains a carbon-oxygen white dwarf, which is key, for example, to the Type Ia 
supernovae. At the same time, the envelope of an AGB
star is less tightly bound than that of a RG, where an envelope
removal due to transfer of orbital energy of the cores to the gas
during the plunge-in phase of CEE is not achieved. We investigate whether this change in
the case of AGB primaries and what determines the final orbital
separation of the stellar cores in such scenarios.

These questions remain unanswered. Only a few three-dimensional
simulations of CEE with AGB primaries have been published, and, because of the numerical difficulties,
low-mass AGB stars have largely been avoided. The restricted computational
resources available at the time limited the numerical resolution of
the simulations presented by \citet{sandquist1998a}, who report little
unbound envelope mass. \citet{staff2016b} more recently found about a
quarter of the envelope mass to become unbound in the
interaction. From simulating the first twenty 
orbits of the inspiral, \citet{chamandy2020a} extrapolate that the
envelope of an AGB system might be ejected within ten years. None of these
studies accounted for the release of ionization energy in the
expanding envelope.

Here, we present high-resolution simulations of CEE with a low-mass AGB
primary star that follow the evolution up to over one hundred
orbits. We compare the limiting case in which recombination energy is
assumed to be completely transferred to the envelope gas with models
where this effect is ignored. We confirm earlier results that suggest that 
when ignoring ionization effects, a larger fraction of the envelope mass is
ejected in CEE with an AGB primary compared to the case of an RG
primary. When accounting for recombination energy release, most of the
envelope is ejected during our simulations and complete envelope
removal is possible. In a series of simulations, we study the impact
of the mass ratio between the primary and the secondary star on the
evolution. Based on our results, we discuss the energy formalism that
is often employed to parameterize CEE with AGB primaries. 

The structure of the paper is as follows: We introduce the physical
model and numerical methods in Sect.~\ref{sec:methods}. Results from
hydrodynamical simulations are presented in Sect.~\ref{sec:cesim12}.
After a discussion in Sect.~\ref{sec:discussion}, we conclude in
Sect.~\ref{sec:conclusions}.

\section{Methods}
\label{sec:methods}

The simulations in this paper are carried out with the hydrodynamics
code \arepo \citep[][]{springel2010a}, which we briefly describe in
Sect.~\ref{sec:method1}.  The initial model for the \arepo simulation
is created with the stellar-evolution code \mesa
\citep[][Sect.~\ref{sec:method3}]{paxton2011a,paxton2013a,paxton2015a}.
Because of the large sound speed and hence short dynamical time scale
in the central region of the star, the stellar core is removed from
our simulations (Sect.~\ref{sec:method4}). Discretization
uncertainties lead to an initial deviation from hydrostatic
equilibrium, which is why we relax the initial model on the \arepo
grid (Sect.~\ref{sec:relaxresults}) before carrying out CE
simulations. The resulting binary set-up is presented in
Sect.~\ref{sec:binset}.

\subsection{Moving-mesh hydrodynamics code}
\label{sec:method1}

We use the finite volume hydrodynamics code \arepo
\citep{springel2010a} with a moving unstructured mesh based on a
Voronoi tessellation of space. The Euler equations are solved on this
mesh with a finite volume approach, based on a second-order unsplit
Godunov scheme. In principle, the motion of mesh-generating points can be 
arbitrary. Setting their velocities to the fluid velocities in the
corresponding cells (plus a regularization component) leads to a nearly 
Lagrangian scheme for which the truncation error is Galilean
invariant. By adaptively refining the mesh, the resolution can be
adjusted according to predefined criteria. The main criterion 
for refinement is to enforce a constant mass of the cells. 
Self-gravity of the stellar-envelope gas cells is included with a tree-based algorithm. 
In our simulations, however, we make use of \arepo's capability to 
include a different kind of particles: point masses that interact only 
via gravitation. These are used to represent the core of the primary star 
and the entire companion star. Although the latter may not be a bare 
stellar core, we refer to these particles as ``core particles'' in the following. All
gravitational interactions involving them are not treated with a tree-based approximation, but exactly. 
Magnetic fields are not considered in this work. 

We compare simulations that assume an ideal-gas equation of state
(\eos) without radiation pressure and the tabulated OPAL \eos \citep{rogers1996a,rogers2002a},
which also accounts for the ionization state of the gas.  Ionization
effects are thought to play an important role in CE ejection, because
the plasma recombines when the material expands and cools down. How
much of the recombination energy contributes to envelope ejection
depends on whether it thermalized locally and ultimately converted
into kinetic energy of the envelope gas or, instead, released in
optically thin regions close to the photosphere and radiated away. In
our models, we do not consider radiation transport, which implies that
released ionization simply adds to the thermal energy at the place
where recombination happens. This treatment potentially overestimates the amount of recombination energy that is
absorbed by the envelope. Our simulation with an ideal-gas \eos tests the opposite limiting case of no recombination energy contributing to envelope removal.

\subsection{Initial stellar model}
\label{sec:method3}

We use the one-dimensional stellar-evolution code \mesa
\citep{paxton2011a,paxton2013a,paxton2015a} in version 7624 to evolve
a $1.2\,M_{\odot}$ zero-age main sequence (ZAMS) mass star with metallicity $Z = 0.02$ to the
AGB stage. Otherwise, we use default settings, for instance, a mixing-length parameter alpha of 2.0. 
As in \citet{paxton2013a}, stellar wind mass loss is included via the Reimers
prescription \citep{reimers1975a} with $\eta = 0.5$ for RG winds and
the Bl{\"o}cker prescription \citep{bloecker1995a} with $\eta = 0.1$ for
AGB winds. 

This model for the primary star in the subsequent CEE simulations is evolved until its radius on the AGB exceeds that at the tip of the RGB such that a CE phase would not have occurred already during the RGB evolution. By the time we end the stellar evolution calculation on the AGB  at an age of $\SI{6.3e9}{\year}$, the original $1.2\,M_{\odot}$ ZAMS model has reached a radius of $R = 173\,R_{\odot}$ and its remaining mass is $M_1 = 0.97\,M_{\odot}$ because some material is lost in stellar winds. 
The model is an AGB star undergoing He-shell burning and there have been no thermal pulses yet. 
Throughout this paper, we refer to this primary star by the mass $M_1$ it has at the onset of the CEE.

\subsection{Set-up of initial \arepo model}
\label{sec:method4}

The \mesa model of the primary star is mapped onto the grid of the \arepo 
code. To avoid the restrictively small numerical time steps required in the dense 
material of the stellar core, we remove the central region up to a 
cut-off radius, $r_{cut} = 0.05 R = 8.7\,R_{\odot}$, and replace it by 
a core particle of mass $M_c = 0.545\,M_{\odot}$. The envelope mass of the 
primary star model is thus $M_e = 0.425\,M_{\odot}$.

In order to obtain a stable, hydrostatic envelope structure (hydrostatic equilibrium for $ \rho \vec{g} - \nabla p  = 0$), we use a modified Lane-Emden equation with an additional term accounting for the core, following \citet{ohlmann2017a}. Around the central core particle, mass is assigned to spherical shells with a HEALPix distribution \citep{gorski2005a}. We map density, internal energy, and chemical composition. Unlike the gas of the envelope, the core particles interact only gravitationally. Therefore, they do not have internal energy in our simulations, that is to say we implicitly assume that the internal energy of the stellar core does not change during the CEE.
We integrate density, thus ensuring that the mechanical profile (\ie, its density and pressure structure) is unaltered. Other quantities, however, may differ between the original and the mapped models. In particular, for the ideal-gas \eos, the thermal structure is not preserved and convection properties change (see Sect.~\ref{sec:relaxresults}). 

For achieving stability in the ideal-gas model, a lower resolution with $N=\SI{3e6}{}$ hydrodynamic cells is sufficient while $N=\SI{6.75e6}{}$ cells are necessary in the OPAL-\eos model. Cells are larger at larger radii since densities are lower. 

About 1.6\% of the star's envelope mass ($0.02\,M_{\odot}$) is assigned to the core particle. This is caused by cutting off at 5\,\% of the stellar radius below which there is still mass belonging to the envelope. The resulting difference in the binding energy of the envelope compared to the \mesa model is about 0.7\,\%. 

The total energy of the AGB \arepo models including internal energy is $\SI{-1.5e46}{\erg}$ with the ideal gas and $\SI{-0.3e46}{\erg}$ with the OPAL \eos (including $\SI{1.2e46}{\erg}$ recombination energy). Both model stars are bound, whereby the envelope of the ideal-gas model is more tightly bound than in the OPAL \eos case.

\subsection{Relaxation}
\label{sec:relaxresults}

The mapping procedure does not guarantee a stellar structure that is in 
hydrostatic equilibrium. To obtain a stable configuration, a 
relaxation following \citet{ohlmann2017a} is performed for ten
acoustic timescales of the envelope ($\SI{720}{\day}$). This 
ensures that flows in the subsequent binary simulations develop only from the 
interaction with the companion star and are not a result of an 
unstable setup. Velocities are damped for half the duration of the
relaxation run, the rest of the time is used for verification of
stability. During the run, cells are refined and de-refined in order
to keep the mass per cell approximately evenly distributed.

Simulations with AGB stars so far encountered problems setting up a
stable envelope \citep[see][]{ohlmann2017a}. Simply increasing the
global resolution does not guarantee stability. Instead, \citet{ohlmann2016a} showed the
importance of resolving the region around the core particle within 
the cut-off radius with a certain
number of cells in order to construct stable models. For ensuring
hydrostatic equilibrium, the pressure gradient that balances gravity
has to be resolved, which requires a certain minimum number of cells
in the radial direction. In the simulations of \citet{ohlmann2016a},
20~cells within the cut-off radius $r_{cut}$ were
sufficient to obtain a stable model of an RGB star in hydrostatic
equilibrium. 
However, because of a steeper pressure gradient around the core in our
AGB model, this resolution is insufficient. A resolution of 40~cells
per cut-off radius is needed in our case to obtain a stable model.

\begin{figure}
    \centering
    \includegraphics[width=\columnwidth]{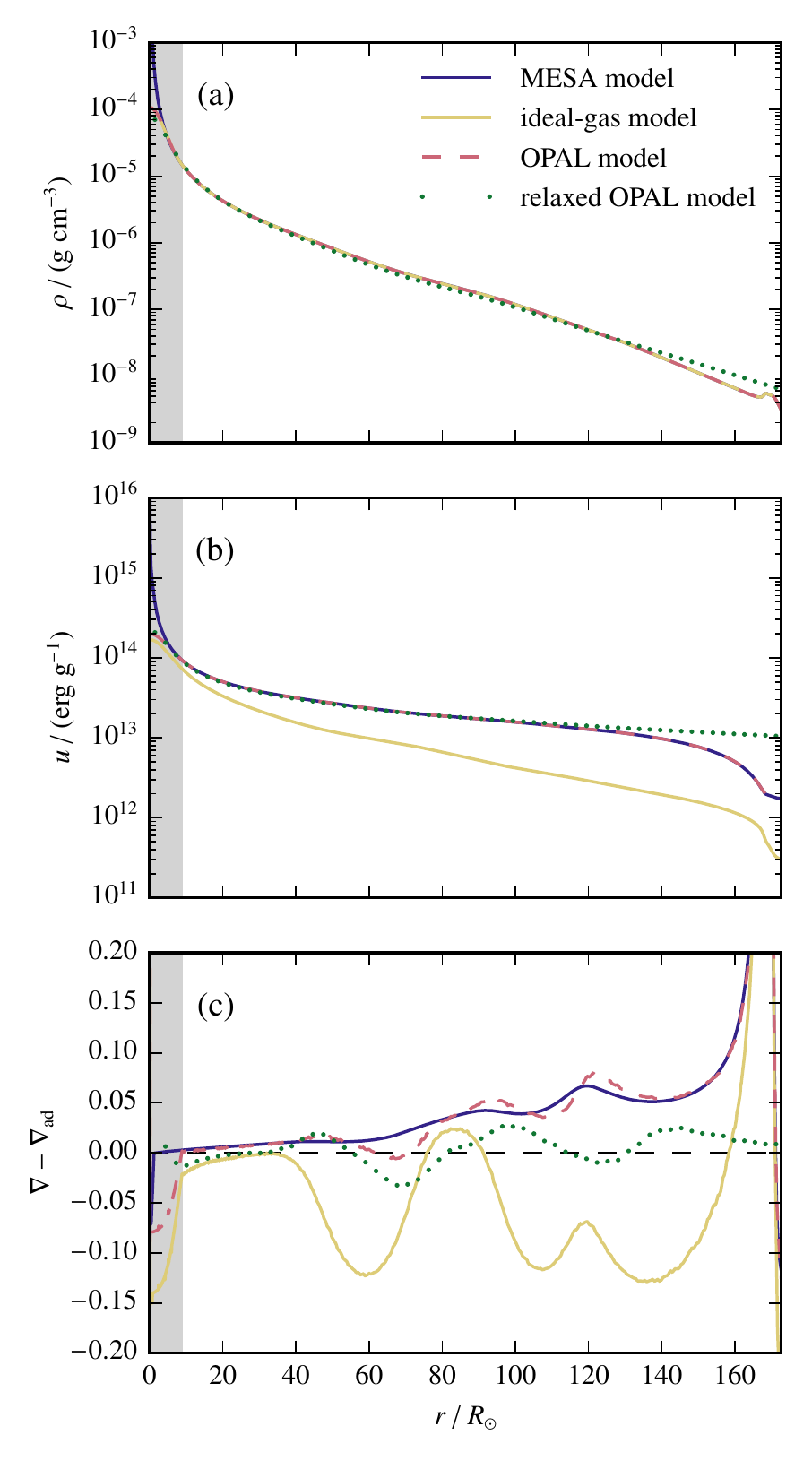}
    \caption[Initial profiles of density and internal
      energy.]{Comparison of the initial profiles of the initial \mesa model for the primary star (blue)
      and the modified models with ideal-gas (yellow) and OPAL (red) equations of state before the relaxation. 
      The gray area indicates the region inside the cut-off radius $r_{\mathrm{cut}}$. The density $\rho$ is shown in the top 
      panel (a) (the density profile of the \mesa model is hidden behind that of the ideal-gas and OPAL model). The green dots are 
      averaged values of cells in the OPAL model after the relaxation. The
      specific internal energy $u$ is shown in the middle panel (b). 
      The difference between the temperature gradient $\nabla$ and the adiabatic gradient $\nabla_{ad}$ is shown in the bottom panel (c).}
    \label{fig:ic-pres-rho}
\end{figure}

The structure of the primary AGB star after relaxation is illustrated
in Fig.~\ref{fig:ic-pres-rho}. The density of the original \mesa model
is well-reproduced down to the cut-off radius. Inside $r_{\mathrm{cut}}$, the profiles deviate because the core particle replaces some of the material. The internal energy of
the ideal-gas model is reduced by the ionization energy. Therefore it
deviates from the original \mesa model, that includes this energy
component. Our OPAL \eos-based model, in contrast, matches the
internal energy structure very well.

With a resolution of 40~cells per cut-off radius, the maximum Mach numbers inside the star are about $1.0$ and $0.1$ for the OPAL and ideal-gas \eos, respectively.
The deviations from the initial density (and pressure) distribution are small inside the relevant part of the model star's envelope; the material expanding beyond the initial radius comprises about $\SI{1}{\percent}$ of the stellar mass. 
The mass-averaged mean difference of both sides of the hydrostatic equilibrium equation ($| \rho \vec{g} - \nabla p | / \mathrm{max}(|\rho \vec{g}|, |\nabla p|)$) is about $\SI{2}{\percent}$ for the OPAL model and about $\SI{1}{\percent}$ for the ideal-gas model. 
The deviations of the total energy during the relaxation are on the order of a few percent with a maximum in the potential energy of the OPAL model of 3\%. 

\begin{figure}
    \centering
    \includegraphics[width=\columnwidth]{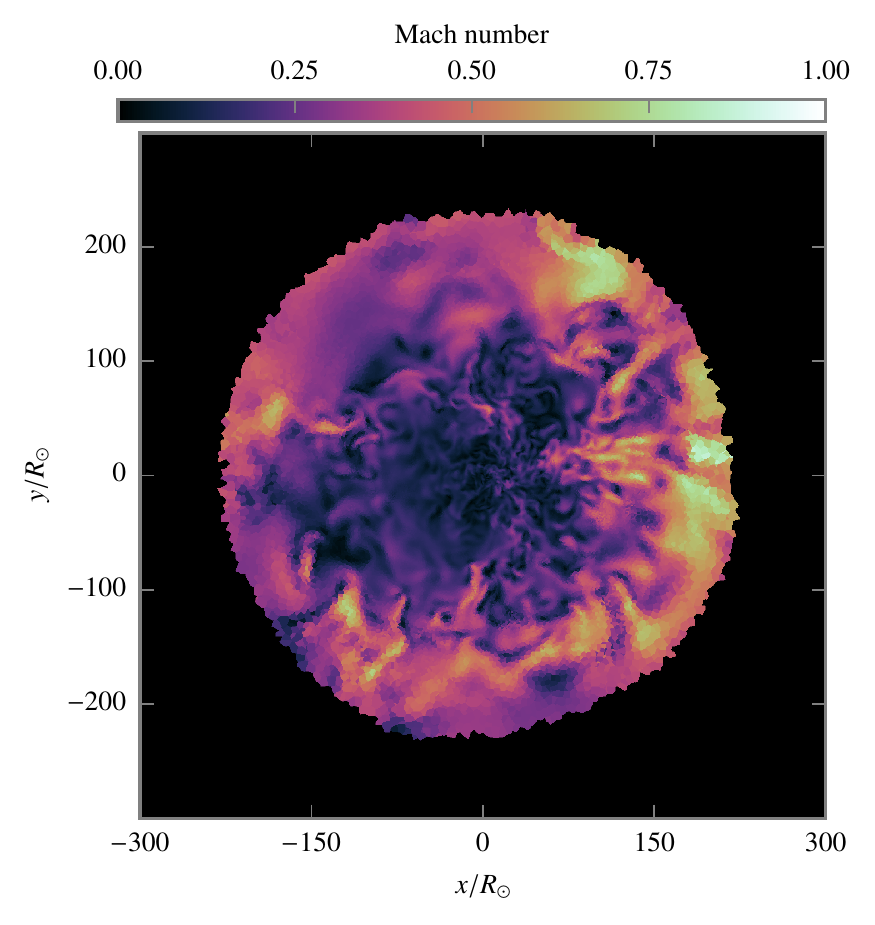}
    \caption[Mach plot of convection during the relaxation.]{Mach number of the flow at the end of the relaxation of the OPAL model. 
      Cells are larger at larger radii since densities are lower. The region outside the star ($\rho < \SI{e-9}{g.cm^{-3}}$) is blacked out.}
    \label{fig:conv-relax}
\end{figure}

The envelope of the giant is convectively unstable in the \mesa model. Mass-averaged mean Mach numbers on the grid after mapping and relaxation are about $0.01$ for the ideal-gas models and $0.10$ for the OPAL models. These flows are attributed to convection. 
We reconstruct the mechanical model with its density and pressure profiles. With the same \eos, the thermal structure of the original \mesa model is retained, but when changing the \eos, it is not necessarily reproduced. This behavior is illustrated by the superadiabaticity $\nabla - \nabla_{\mathrm{ad}}$ in Fig.~\ref{fig:ic-pres-rho}c. For the model with the ideal-gas \eos, the mean deviation in the mapped temperature gradient from that of an adiabatic stratification, 
$(\nabla - \nabla_{ad}) \sim -0.05$, is negative in the stellar interior. Thus, we do not expect convection. 
In contrast, in the model with the OPAL \eos, we obtain $(\nabla - \nabla_{ad}) \sim 0$. Convection is expected and can indeed be observed to develop in our relaxation simulations (Fig.~\ref{fig:conv-relax}).

We calculate the structural parameter $\lambda$ \citep{webbink1984a,dekool1987b} of the AGB star models.
The binding energy can be approximated by
\begin{equation}
E_{bin} = \int_{M_c}^{M_1} \left( -G \frac{M(r)}{r} + \alpha_{th} u \right) \mathrm{d}m \approx - G \frac{M_e \left( M_e + M_c \right) }{\lambda R},
\label{eq:ebinlambda}
\end{equation}
where the total mass of the giant star $M_1$ equals the sum of the core and envelope masses, $M_1 = M_c + M_e$. 
In the integral in Eq.~(\ref{eq:ebinlambda}), we are following \citet{dewi2000a} when including the fraction $\alpha_{th}$ \citep{han1995a} of the internal energy.
Using the right-hand side of Eq.~(\ref{eq:ebinlambda}), we calculate two different values of $\lambda$: $\lambda_g$ for $\alpha_{th}=0$ and $\lambda_b$ for $\alpha_{th}=1$ (Table~\ref{table:lambda}).
Without the internal energy, $\lambda_g \approx 0.32$ for the OPAL \eos and $\lambda_g \approx 0.31$ for the ideal-gas \eos. 
This is consistent with \citet{demarco2011a}, whose fit gives $\lambda = 0.28 \pm 0.03$ for a $1.0\,M_{\odot}$ AGB star. 
An additional factor of $1/2$ in their definition of $\lambda$, however, leads to a systematic offset of $\SI{-22}{\percent}$ compared to our values of $\lambda$. 
Including the internal energy, the envelope is less tightly bound, $\lambda_b \approx 2.83$ for the OPAL \eos and $\lambda_b \approx 0.60$ for the ideal-gas \eos.

\begin{table}[t]
\centering
\caption[$\lambda$ parameter for $1.0\,M_{\odot}$ primary.]{Envelope structure parameter $\lambda$ for the $1.0\,M_{\odot}$ primary.} 
\begin{tabular}{l c c c c}
\toprule
\eos & $E_g\,[\SI{e46}{\erg}]$ & $\lambda_g$  & $E_b\,[\SI{e46}{\erg}]$ & $\lambda_b$\\
\midrule
ideal gas & $\SI{-2.91}{}$ & 0.31 & $\SI{-1.51}{}$ & 0.60 \\
OPAL & $\SI{-2.86}{}$ & 0.32 & $\SI{-0.32}{}$ & 2.83 \\
\bottomrule
\end{tabular}
\label{table:lambda}
\end{table}

\subsection{Binary setup}
\label{sec:binset}
We use the initial stellar models discussed above to set up our binary-evolution simulations of the hydrodynamics of CE phases. The relaxed model 
is placed into a box large enough for the expelled matter not to leave the simulation domain during the runtime (a box size of $\SI{1.1e5}{R_\odot}$ is chosen). 
The AGB primary star model is placed into a binary system with a less massive companion star of mass $M_2$, which could be a main-sequence star or a white dwarf. Here, we consider mass ratios $q = M_2/M_1$ of 0.25, 0.5 and 0.75. The companion star is represented by a point mass. To properly resolve the region around point masses (\ie, the core of the AGB star and the secondary star), we prescribe a spatial resolution of 40~cells per cut-off radius (Sect.~\ref{sec:relaxresults}). 
The gravitational force of the gas particles and point masses is smoothed at a length of $h \approx \SI{1.6e-2}{R_\odot}$ and $h \approx 3.1\,R_{\odot}$, respectively, following the spline function given in \citet{springel2010a}. 

The Roche-lobe radius $R_L$ of a binary-star system of mass ratio $q$ and orbital separation $a$ is approximately \citep{eggleton1983a}
\begin{equation}
\frac{R_L}{a} = \frac{0.49 q^{-2/3}}{0.6 q^{-2/3} + \ln (1 + q^{-1/3})} \equiv r_L.
\label{eq:roche-radius}
\end{equation}
Mass transfer starts when the primary AGB star fills its Roche lobe. Subsequently, the orbital separation is expected to shrink and the primary's radius to increase. 
Since we cannot follow this initially slow process in our hydrodynamic simulations, we artificially reduce the initial orbital separation, $a_i$, to 
60\,\% of the separation for which the AGB star would overflow its Roche lobe (\ie, $a_i = 0.6\,R/r_L$), assuming $R_L \approx R$, the primary star's radius. 

In the resulting binary configuration, the companion star is well above the surface of the primary (\cf Table~\ref{table:Lmodels}). 
The companion is set up rotating with the orbital period given by Kepler's law, $P = 2 \pi\, [G(M_1+M_2)/a_i^3]^{-1/2}$. 
We set up solid body rotation of the AGB star envelope with $\SI{95}{\percent}$ of the angular frequency of co-rotation with the companion to facilitate the inspiral. 

The initial orbital separations of the CE simulations with the different mass ratios depend on the mass ratio $q$. This has little effect on the initial orbital energy $E_{orb}$ and the amount of orbital energy used to eject envelope material, $\Delta E_{orb}$, is mostly determined by the final orbital separation, $a_f$.
A list of all models presented in this paper and their respective parameters is given in Table~\ref{table:Lmodels}.

\begin{table}[ht]
\centering
\caption[Setups for binary runs.]{Setups for binary runs with mean number of cells $N$, mass ratio $q$, initial orbital separations $a_i$ and initial orbital periods $P$. The difference in the latter for the same value of $q$ but a different \eos is due to rounding errors.} 
\begin{tabular}{l c c c c c}
\toprule
Model & \eos & $N$ & $q$ & $a_i\,[R_{\odot}]$ & $P\,[\SI{}{d}]$\\
\midrule
I.25 & ideal gas & $\SI{2e6}{}$ & 1/4 & $\SI{207}{}$ & 312\\
I.50 & ideal gas & $\SI{2e6}{}$ & 1/2 & $\SI{236}{}$ & 347\\
I.75 & ideal gas & $\SI{2e6}{}$ & 3/4 & $\SI{257}{}$ & 365\\
O.25 & OPAL & $\SI{6.75e6}{}$ & 1/4 & $\SI{207}{}$ & 312\\
O.50 & OPAL & $\SI{6.75e6}{}$ & 1/2 & $\SI{236}{}$ & 346\\
O.75 & OPAL & $\SI{6.75e6}{}$ & 3/4 & $\SI{257}{}$ & 364\\
\bottomrule
\end{tabular}
\label{table:Lmodels}
\end{table}

\section{Common-envelope simulations}
\label{sec:cesim12}

We present the results of the binary simulations:
In Sect.~\ref{sec:ref-runs}, we start by comparing in detail two reference runs with $q=0.5$ for the ideal-gas and the OPAL \eos. 
Mass unbinding is analyzed in Sect.~\ref{sec:mass-unbind} and recombination-energy usage in Sect.~\ref{sec:erecomb-use}. 
We discuss the results for different companion masses in Sect.~\ref{sec:diff-comp}, the final-to-initial-separation ratio in
Sect.~\ref{sec:fin-sep}, and the $\alpha$-formalism in Sect.~\ref{sec:alpha-res}.

\subsection{Reference runs}
\label{sec:ref-runs}

\begin{figure}
  \centering
  \includegraphics[width=\columnwidth]{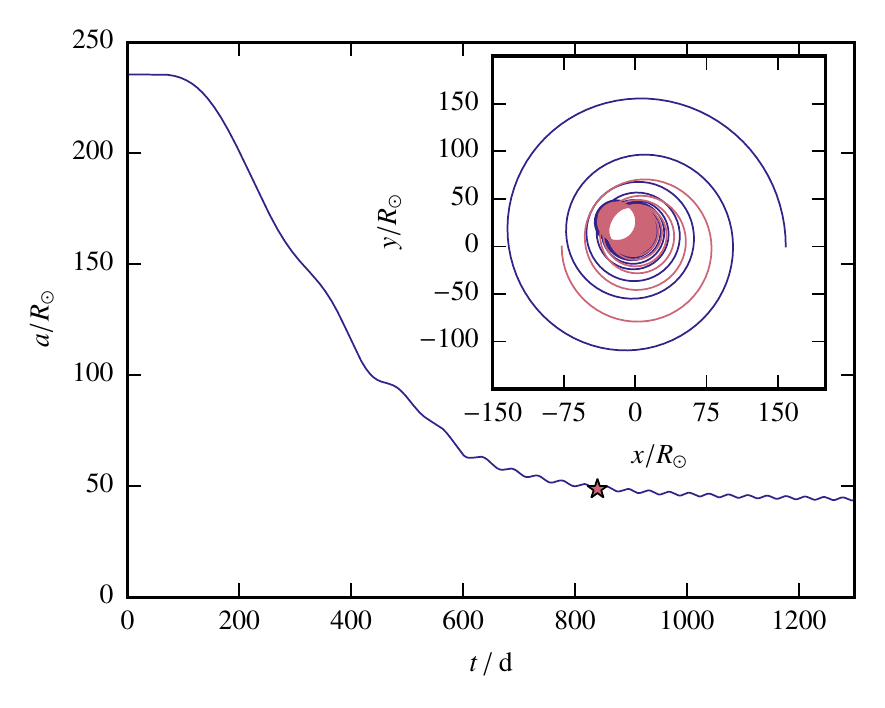}
  \caption[Orbital evolution.]{Orbital evolution of the O.50 reference run. The oscillations are due to eccentricity in the inspiral, the star symbol marks the end of the plunge-in phase. In the inset we show the trajectory of the primary in red and that of the secondary star in blue.}
  \label{orbit-prior-opal}
\end{figure}

A first CEE simulation with $q = 0.5$ is conducted for both the OPAL-\eos (O.50) and the ideal-gas (I.50) models. 
The total energy of the systems, defined as the sum of the potential, kinetic and internal energies of the cores (respective primary star core and companion) and the envelopes, is still negative: $\SI{-0.6e46}{\erg}$ for the OPAL-\eos and $\SI{-1.8e46}{\erg}$ for the ideal-gas model. Therefore, the envelopes of both systems can become entirely unbound only by the additional release of orbital energy. 
We run the O.50 (I.50) model for about $\SI{2500}{\day}$ ($\SI{2000}{\day}$), amounting to $\SI{1.3e6}{}$ ($\SI{1.1e6}{}$) time-step integrations. We follow about 60.4 (70.0) binary orbits, of which 51.2 (56.1) are after the end of the plunge-in, where we define the plunge-in phase as $|\dot{a} P|/a > 0.01$, in contrast to \citet{ivanova2016a} who require $|\dot{a} P|/a \gtrsim 0.1$). In Fig.~\ref{orbit-prior-opal}, we show the orbital evolution of model O.50. 

\begin{figure*}[ht]
\centering
\includegraphics[width=1.0\textwidth]{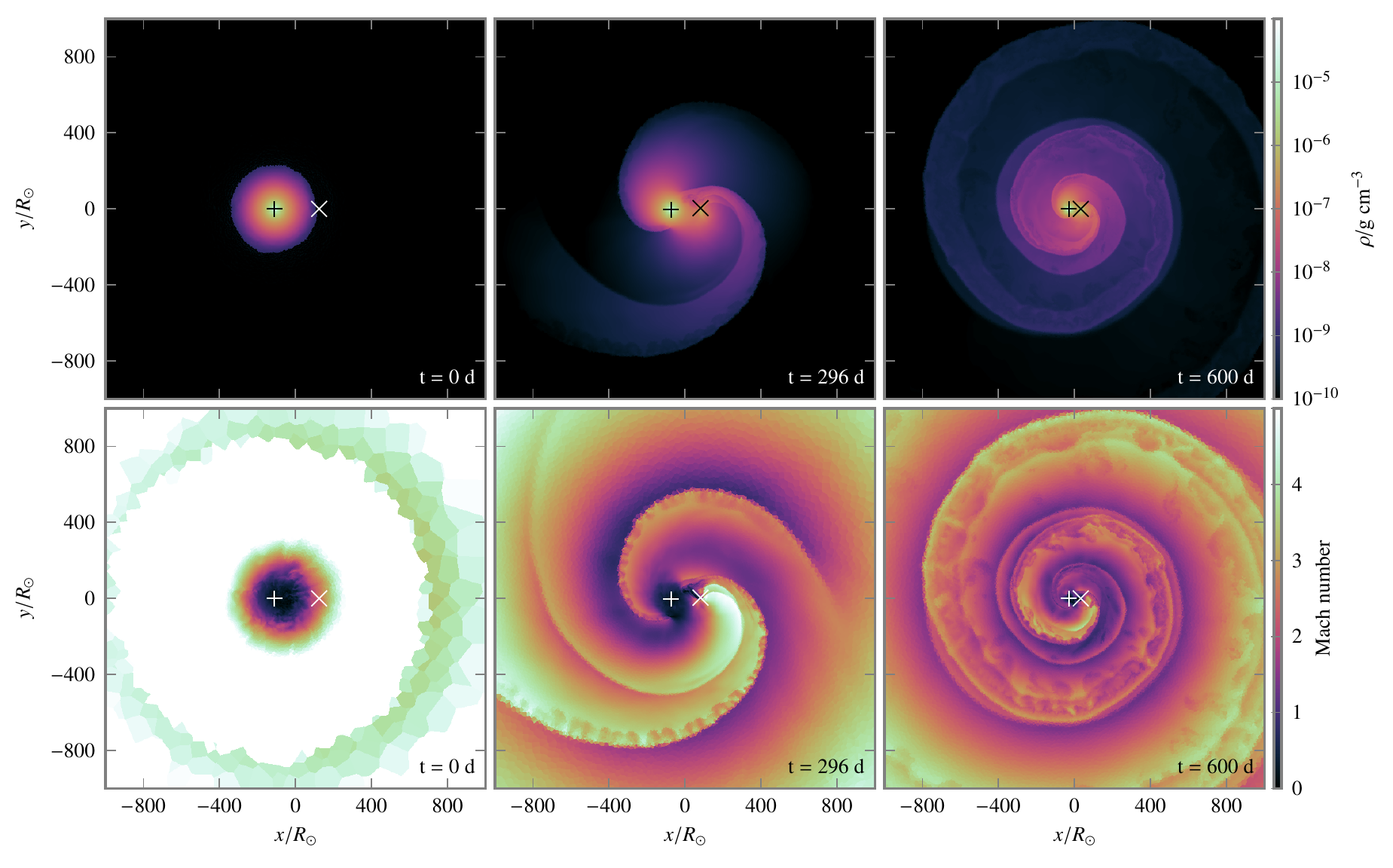}
\caption[Time series of density and Mach plots.]{Density and Mach number of model O.50 in the orbital plane at the beginning of the simulation, after one and after four orbits. The core of the primary star is marked by a plus symbol ($+$) and the secondary star by a cross symbol ($\times$). 
The regions between the supersonic shocks are transonic, outside of the spiral structure it is supersonic due to the rotation in the setup, the low density and the pressure.}
\label{fig:12binopalmach}
\end{figure*}

In Fig.~\ref{fig:12binopalmach}, we show the evolution of density and Mach number of the OPAL reference run in the orbital ($x$-$y$) plane (similar evolution for the ideal-gas \eos). 
The O.50 (I.50) run starts with an initial orbital period of $\SI{346}{\day}$ ($\SI{347}{\day}$) in the left panels of Fig.~\ref{fig:12binopalmach}; the dynamical timescale of the envelope is $\SI{30}{\day}$ and the acoustic timescale is $\SI{72}{\day}$. 
The secondary star is enclosed in the common envelope after about $\SI{68}{\day}$. 
The cores are surrounded by overdense material and compress material by moving through the now common envelope, forming two spiral arms moving outward. 
The orbital frequency increases during the inspiral, and, after one orbit, shear instabilities arise within the spiral arms (middle panels of Fig.~\ref{fig:12binopalmach}); best visible in the Mach number plot in the lower row). The spiral arms themselves also move outward with higher velocities than previously so that subsequent shock layers hit previous layers. The second arm collides with the first after about $1.6$ orbits ($\SI{400}{\day}$) at a distance of about $700\,R_{\odot}$ from the center of mass. 
After four orbits ($\SI{600}{\day}$, right panels in Fig.~\ref{fig:12binopalmach}), the radius at which consecutive layers hit each other has shrunk to about $200\,R_{\odot}$ and instabilities form inside this radius where the density is decreased due to the expansion of the AGB primary star.

\begin{figure}[ht]
\centering
\includegraphics[width=\columnwidth]{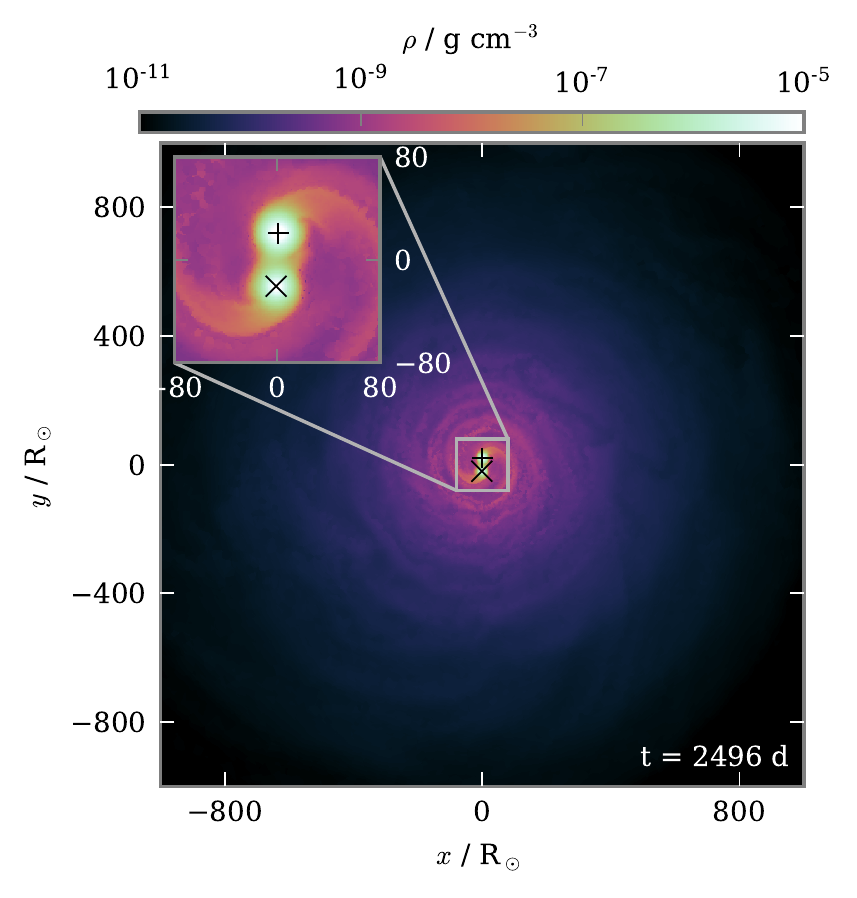}
\caption[Late spiral structure.]{Density in the orbital plane of O.50 at the end of the run. The preserved spiral structure from the inspiral is visible. The core of the primary is marked by a plus symbol ($+$) and that of the secondary by a cross symbol ($\times$).}
\label{fig:12binopallate}
\end{figure}

In the case of the O.50 (I.50) run, $\SI{840}{\day}$ ($\SI{766}{\day}$) after the start of the plunge-in, 
this phase is terminated at an orbital separation of $50\,R_{\odot}$ ($41\,R_{\odot}$) and period $\SI{39}{\day}$ ($\SI{29}{\day}$). 
The slower inspiral in case of the OPAL-\eos run is probably due to the expansion of the envelope accelerated by the release of recombination energy. 
At the end of the run, after $\SI{2500}{\day}$ ($\SI{2000}{\day}$), the remaining orbital separation has reduced to $41\,R_{\odot}$ ($34\,R_{\odot}$) and the period is $\SI{31}{\day}$ ($\SI{23}{\day}$). The final orbit has a low eccentricity of $e \approx 0.038$ ($\approx 0.010$). We measure the eccentricity with the technique proposed by \cite{ohlmann2016a} by fitting ellipses.
At the end of the ideal-gas run, the mass ejection has stagnated; in case of the OPAL \eos, it is still going on. 
Although increasingly affected by the instabilities, the spiral pattern can still be recognized at late times (Fig.~\ref{fig:12binopallate}), where -- as visible in the inset -- the two cores still create new shock waves and maintain the spiral structure.
We summarize the orbital evolution of the ideal-gas and OPAL-\eos runs for the total run and the plunge-in phase separately in Table~\ref{table:Ldyncomp}. 

The angular momentum of the cores as well as their kinetic energy are continuously transferred to the envelope during the plunge-in phase. 
The transfer of energy of the cores to the envelope mainly acts via the (tidal) drag force in close vicinity of the cores. 
The error on the angular momentum increases with simulation time to a maximum of $\SI{1.1}{\percent}$ ($\SI{1.3}{\percent}$) at the end of the simulation using the OPAL (ideal-gas) \eos.

\subsection{Unbound mass}
\label{sec:mass-unbind}

Orbital energy transferred from the cores to the envelope is only one contribution to the overall process powering mass ejection, in particular in case of the OPAL \eos. 
In fact, it is chiefly the thermal and internal energies of the envelope itself that are converted into kinetic energy (see Table~\ref{table:Lcompeal}). This becomes evident in particular after the plunge-in, when the orbital separation approaches its final value in the simulations.
The error in the total energy at the end of plunge-in phase is only 2.0\,\%. 
Unfortunately, after $\SI{2500}{\day}$ ($\Delta E_{tot} = \SI{1.91e45}{\erg}$, equivalent to a gain of 5.3\,\% of the initial potential energy), the energy error rate of $\SI{1.0e37}{\erg \s^{-1}}$ is larger than the recombination-energy-release rate in the O.50 run and we can no longer reasonably decide whether material becomes unbound or not. We thus choose $t = \SI{2500}{\day}$ as the end of the reference run. We finish the I.50 run after $\SI{2000}{\day}$ with $\Delta E_{tot} = \SI{0.76e45}{\erg}$, that is a gain of 2.1\,\% of the initial potential energy. 

The unbound mass is defined as the sum of the masses of cells having a positive energy. It is not readily clear how different forms of energy can contribute to the unbinding and we thus consider different criteria (Table~\ref{table:Lcompeal}): 
$f_{ej,kin}$ is the fraction of the envelope mass unbound when only accounting for kinetic energy to balance gravitational binding energy, $f_{ej,therm}$ is the fraction unbound when comparing the potential energy of the material to the sum of its kinetic and thermal energies, and $f_{ej,OPAL}$ is the fraction unbound when accounting for kinetic plus total internal (thermal and recombination) energy in case of the OPAL \eos. 
The save conservative definition is to regard as unbound only material for which the kinetic energy exceeds its gravitational potential energy. Ultimately, however, thermal energy will be converted into kinetic energy and also recombination energy may increase it. 
Because the total energy error in the simulations exceeds the recombination-energy release at late times, we cannot follow the recombination until the unbound mass fraction saturates in our OPAL-\eos based simulations. Instead, we determine the unbound mass fraction by including the recombination energy still stored in the gas in the criterion $f_{ej,OPAL}$. If energy errors could be avoided, this is what our simulations are bound to arrive at. Of course, there is the possibility that energy is lost by radiation before the envelope is completely removed. This effect is currently not accounted for in our modeling (see Sect.~\ref{sec:erecomb-use}).

\begin{figure}[ht]
\centering
\includegraphics[width=\columnwidth]{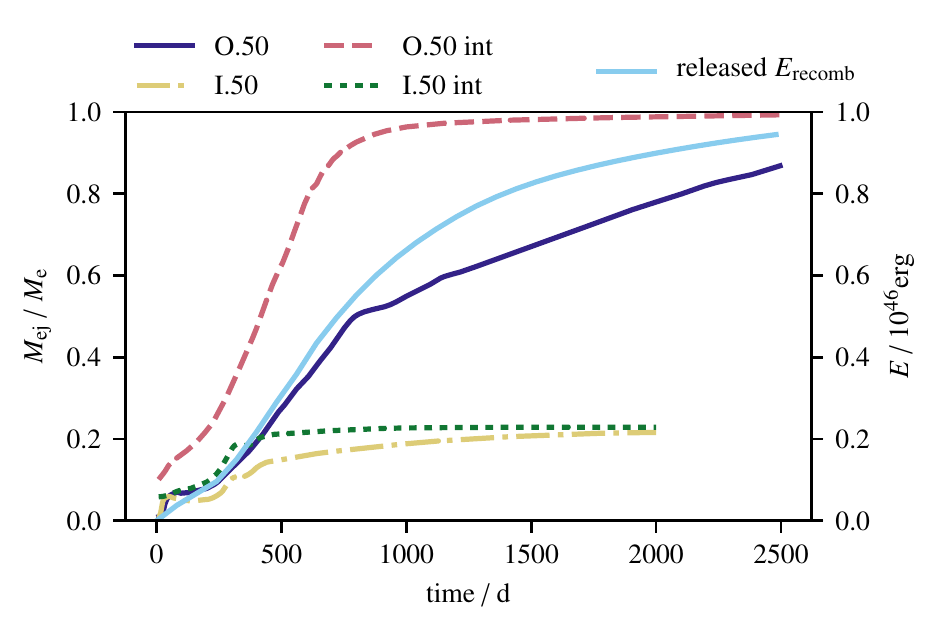}
\caption[Unbound mass evolution.]{Fraction of ejected mass, $M_{ej}$, over time for the O.50 (dark blue and red) and the I.50 (yellow and green line) models. It is plotted with kinetic energy only (dark blue and yellow line) and including internal energy (red and green line), respectively, for I.50 (where internal means thermal energy) and O.50 (where internal means thermal plus recombination energy), respectively. The released recombination energy is shown in light blue on the right-hand axis.} 
\label{massloss-prior-ref}
\end{figure}

For the two reference runs, the unbound mass fractions are plotted versus time in Fig.~\ref{massloss-prior-ref}. For I.50, it saturates at about $\SI{20}{\percent}$ and we can follow the conversion of thermal energy, while in O.50 almost the entire mass is unbound when accounting for the stored internal (ionization) energy. Here, recombination occurs as matter from the envelope cools down when expanding. This leads to further unbinding of material on longer timescales. 
Most of the recombination energy is only released at rather late times, see Fig.~\ref{massloss-prior-ref}. 
Still, not the entire conversion can be followed in our simulation. We reach 86\,\% unbound mass after $\SI{2500}{\day}$ when accounting for kinetic energy alone and the unbound mass according to this criterion is still increasing when we terminate our simulation. 
After the plunge-in, about $\SI{10}{\percent}$ of the envelope's mass is still inside the orbit of the two cores, but the final mass fraction within the orbit of the two cores, $f_{inorb}$, is only about 2\%. Unless there is a change in the overall structure of the remaining envelope, no further dynamical spiral-in is possible due to the low ambient density and the co-rotation of the material leading to a small relative velocity and a vanishing drag force. 
We discuss the possible fate of the system in Sect.~\ref{sec:discussion}. 

\subsection{Recombination energy}
\label{sec:erecomb-use}

We observe similar dynamics for the ideal-gas and OPAL runs, but the ejected mass is larger when including recombination energy. 
Ions recombine once the envelope has cooled and expanded sufficiently. The released recombination energy can lead to further mass unbinding even after the plunge-in phase has terminated. 
Because no radiation transport is implemented in our simulations, all of the released recombination energy is thermalized and absorbed locally. This may overestimate the unbound mass, because some radiation may escape in low opacity regions and would therefore not be available to help unbind mass. 

\begin{figure*}[ht]
\centering
\includegraphics[width=1.0\textwidth]{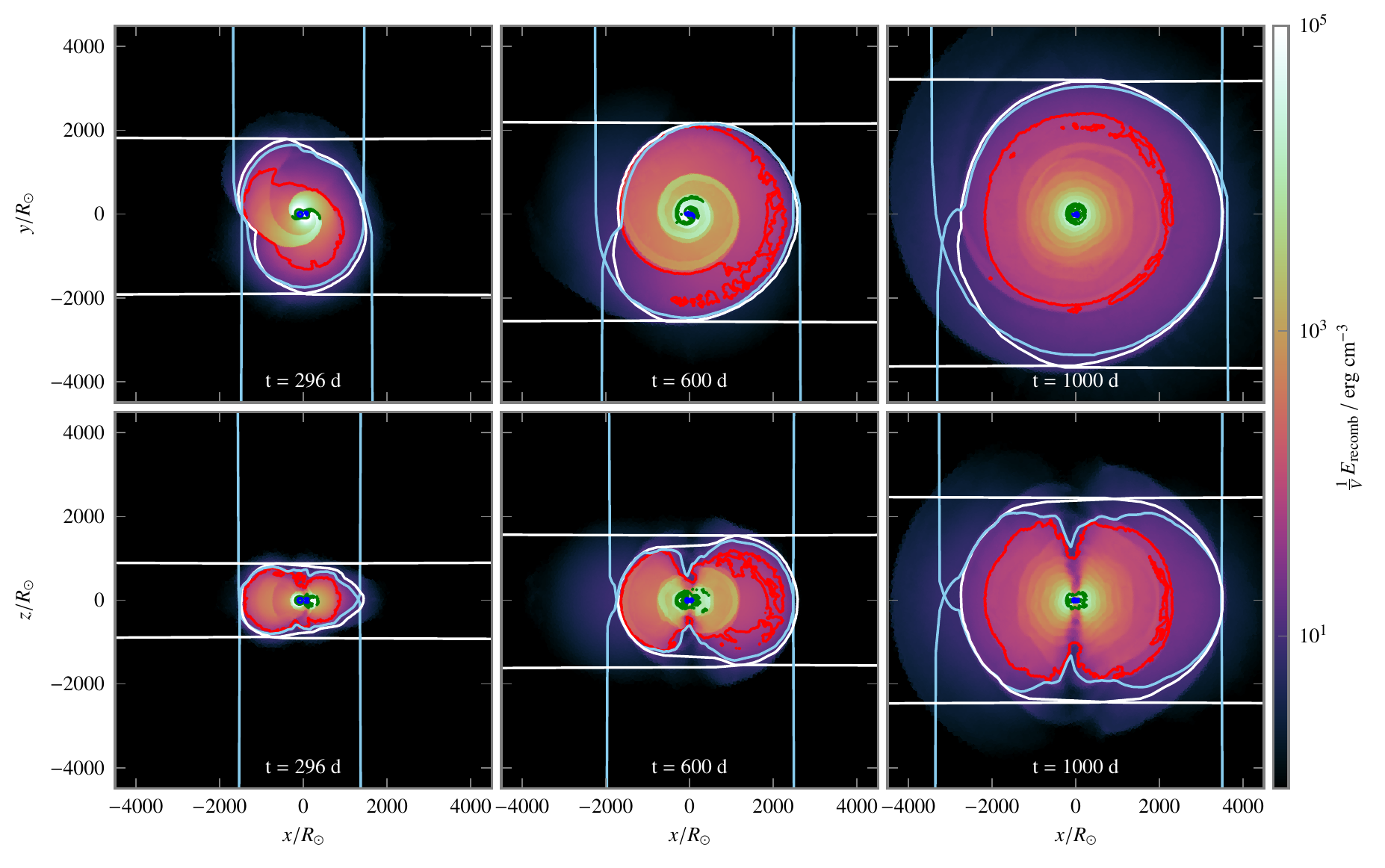}
\caption[Time series of remaining recombination energy plots.]{Available ionization energy of Model O.50 in the orbital plane at the beginning of the simulation, after one and after four orbits. The white and light blue lines represent projections of the photosphere along the axes (\ie, they outline the opaque region). 
The red, green and dark blue contours enclose regions with ionization fractions of H, HeI, and HeII larger than 0.2, respectively.}
\label{fig:12binmfperecomb}
\end{figure*} 

In Fig.~\ref{fig:12binmfperecomb}, we show in a time series the remaining recombination energy and an approximation of the position of the photosphere. The latter is determined by integrating the optical depth $\tau$ along the $x$, $y$ and $z$ directions from outside in until reaching $\tau=1$. 
Regions inside the spiral structure (between the layers) have a lower remaining ionization energy (see Fig.~\ref{fig:12binmfperecomb}), meaning energy has been released by recombination. This indicates that recombination acts behind the spiral shocks where the gas cools, boosting the expansion. 

Regions with an ionization fraction of $> 0.2$ for H, HeI and HeII are indicated by isocontours. The remaining recombination energy is mainly inside the photosphere. The approximated photosphere encloses the region up to an ionization fraction of about 0.1 for hydrogen. This suggests that most of the recombination energy is released inside the photosphere such that it can thermalize quickly and thus be used for unbinding the envelope.
At the times shown in Fig.~\ref{fig:12binmfperecomb}, $\SI{0.9}{\percent}$, $\SI{2.4}{\percent}$ and $\SI{6.1}{\percent}$, respectively, of the recombination energy is outside of the photosphere and thus cannot be completely used for envelope unbinding. It is, however, about $\SI{20}{\percent}$ at the end of the simulation ($\SI{2500}{\day}$). Comparing the initially available recombination energy ($\SI{1.2e46}{\erg}$, Sect.~\ref{sec:method4}) to the recombination energy released by then ($\SI{0.95e46}{\erg}$, Table~\ref{table:Lcompeal}), we note that this share may not be needed to achieve full envelope ejection.

Dust is expected to form at the end of the plunge-in \citep{iaconi2020a}, which would significantly increase the opacity of the material. One requirement for condensation and formation of dust is a temperature below about $\SI{2000}{K}$ \citep{nozawa2013a}.
At the times specified in Fig.~\ref{fig:12binmfperecomb} ($\SI{296}{\day}$, $\SI{600}{\day}$ and $\SI{1000}{\day}$), the innermost cell with a temperature below $\SI{2000}{K}$ is at distance of $1350\,R_{\odot}$, $2350\,R_{\odot}$ and $3875\,R_{\odot}$ to the center, that is about the radius of the photosphere, with densities up to $\SI{2e-13}{\g \per \cubic \cm}$. After $\SI{2500}{\day}$, it is $8500\,R_{\odot}$. 
The condensation radius according to \cite{glanz2018a} for $\SI{1500}{K}$ (effective temperature $T_*^i = \SI{3170}{K}$ for our AGB model) is $586\,R_{\odot}$. After $\SI{2500}{\day}$, about $\SI{6}{\percent}$ of the envelope mass are still bound each inside and outside this condensation radius measured by the kinetic energy criterion. 
Because the background temperature on our grid is $\SI{1870}{K}$, we cannot follow gas down to the $\SI{1500}{K}$ assumed by \cite{glanz2018a}. The condensation radius for $\SI{2000}{K}$, however, is $329\,R_{\odot}$, much smaller than the radius of $8500\,R_{\odot}$ we found after $\SI{2500}{\day}$. 
Even when ignoring dust formation, most of the recombination energy is released in optically thick regions, but potential dust formation further supports our assumption that at least large parts of the recombination energy can be used for envelope ejection at late times. 

\subsection{Different companion stars}
\label{sec:diff-comp}

We now compare CE simulations of binary systems with the same AGB primary star as before, but mass ratios between the two interacting stars of $q =$ 0.25, 0.5 and 0.75. The key parameters of the orbital evolution in the different runs are summarized in Table~\ref{table:Ldyncomp}. 

\begin{table*}[ht]
\centering
\caption[Orbital evolution with different companions.]{Orbital evolution with different companions. The orbital separations $a$, the periods $P$, the times $t$, the numbers of orbits $n$ and the eccentricities $e$ are given for the end of the plunge-in with the subscript ``pi'' and for the end of the run with the subscript ``f''.} 
\begin{tabular}{l c c c c c c c c c c c c}
\toprule
Model & $M_2\,[M_{\odot}]$ & $a_i\,[R_{\odot}]$ & $t_{pi}\,[\SI{}{\day}]$ & $n_{pi}$ & $a_{pi}\,[R_{\odot}]$ & $P_{pi}\,[\SI{}{\day}]$ & $e_{pi}$ & $t_{f}\,[\SI{}{\day}]$ & $n_{f}$ & $a_f\,[R_{\odot}]$ & $P_f\,[\SI{}{\day}]$ & $e_f$\\
\midrule
I.25 & $0.24$ & $207$ & $703$ & 18.7 & $23$ & $14$ & 0.070 & $2000$ & 151.5 & $17$ & $10$ & 0.032\\
I.50 & $0.49$ & $236$ & $766$ & 13.9 & $41$ & $29$ & 0.033 & $2000$ & 70.0 & $34$ & $23$ & 0.010\\
I.75 & $0.73$ & $257$ & $971$ & 12.2 & $65$ & $53$ & 0.078 & $2000$ & 38.5 & $55$ & $41$ & 0.083\\

O.25 & $0.24$ & $207$ & $1046$ & 19.9 & $26$ & $17$ & 0.055 & $2500$ & 124.2 & $22$ & $13$ & 0.008\\
O.50 & $0.49$ & $236$ & $840$ & 9.2 & $50$ & $39$ & 0.059 & $2500$ & 60.4 & $41$ & $31$ & 0.038\\
O.75 & $0.73$ & $257$ & $895$ & 6.6 & $79$ & $73$ & 0.149 & $2500$ & 32.6 & $69$ & $57$ & 0.102\\
\bottomrule
\end{tabular}
\label{table:Ldyncomp}
\end{table*} 

\begin{figure}
\centering
\includegraphics[width=\columnwidth]{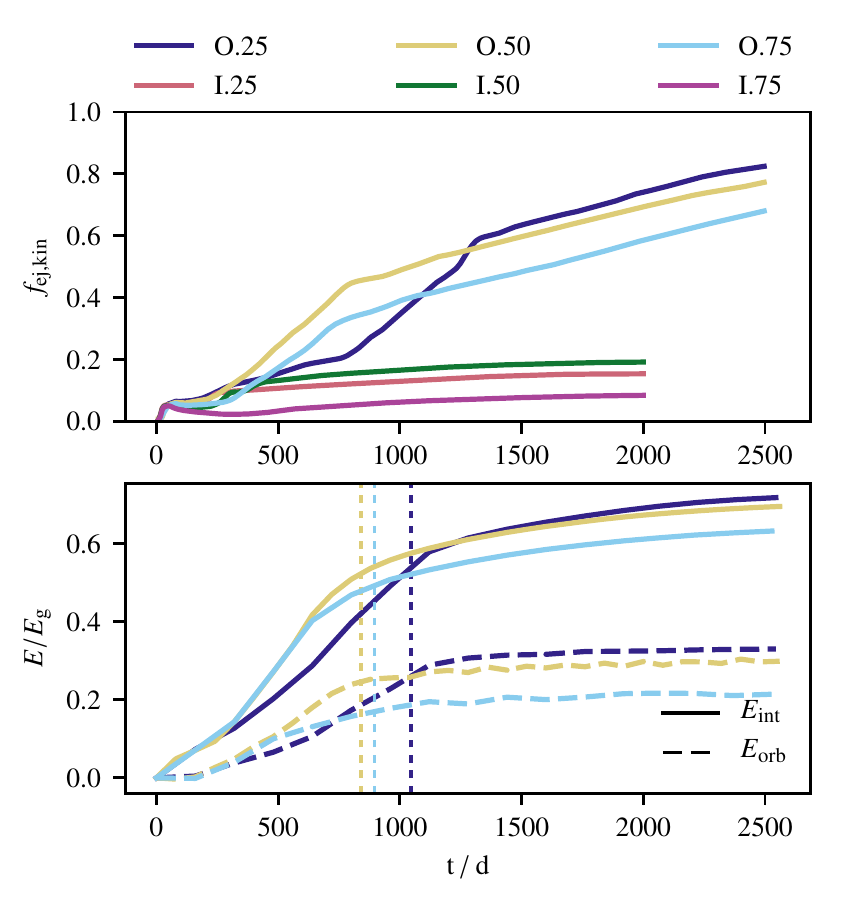}
\caption[Ejected mass and energy budget for different companions.]{Ejected mass and energy budget for different companions. In the upper panel, the unbound mass fraction over time accounting for kinetic energy only is shown. Dark blue, yellow and light blue lines (OPAL \eos), red, green and purple lines (ideal-gas \eos) are for $q=0.25$, $q=0.5$ and $q=0.75$, respectively. 
In the lower panel, the released internal (continuous) and orbital energies (dashed lines) are plotted as fractions of the gravitational binding energy for the OPAL runs.
The vertical dotted lines indicate $t_{pi}$ from Table~\ref{table:Ldyncomp}.}
\label{massloss-prior-companhdideal}
\end{figure}

For the ideal-gas \eos, a more massive companion leads to a slower plunge-in, while there is no such clear trend for the OPAL-\eos runs (see values for $t_{pi}$ in Table~\ref{table:Ldyncomp}). In terms of release of orbital energy, however, the ideal gas and OPAL-\eos runs are qualitatively similar (see Fig.~\ref{massloss-prior-companhdideal}). In case of O.25, the energy release is delayed but higher values from $\SI{1300}{\day}$ on (corresponding to 36.4 [O.25], 22.3 [O.50] and 12.6 [O.75] orbits). Considering the total run, less massive companions spiral in deeper \citep{podsiadlowski2001a} and release more orbital energy (when measured at the same time; see Table~\ref{table:Lcompeal}). This trend is not unexpected: according to \citet{paczynski1976a}, the drag force is weaker for a lower secondary mass, thus transfer of orbital energy to the envelope is less efficient and a deeper inspiral is possible.  
We find that the spiral-in is deeper by 17\,\% to 23\,\% when not including recombination energy compared to the final separations in the simulations that include recombination energy. This can be explained by the fact that without recombination energy release the expansion of the envelope is slower and the transfer of orbital energy terminates later when little mass is within the orbit of the cores. 
The eccentricity of the orbit is larger for more massive companions because the inspiral is quicker; vice-versa, it is smaller for less massive companions where more orbits are required to reach the final separation and thus the motion circularizes. 

In Table~\ref{table:Lcompeal}, we compare the unbound mass fractions for different energy contributions with different companions at the end of the runs. 
For the OPAL \eos, less massive companions can unbind more of the envelope's mass until the end of the run, but all fractions are still increasing at that time (Fig.~\ref{massloss-prior-companhdideal}). For the ideal-gas \eos, the unbound mass fraction has saturated when the simulations are terminated. 
For the OPAL-\eos runs, the release of orbital energy stagnates and the release rate of internal energy decreases at about $t_{pi}$, leading to a change of slope in the curve of the unbound mass fraction (Fig.~\ref{massloss-prior-companhdideal}). 
Our criterion for the plunge-in phase, $|\dot{a} P|/a > 0.01$, captures the instant of slope change accurately for O.25 and O.50; however, the change of slope is not very pronounced in case of O.75 and therefore the criterion is not precise.

The different amounts of released orbital and internal energy due to the slower and deeper inspiral explain the trend in the unbound mass: at the same elapsed physical time, the system with the smallest mass ratio has expelled the largest fraction of envelope material. However, as the mass unbinding still continues by the end of our simulations, this is a statement about the efficiency of mass ejection rather than its overall success. We observe a stronger dynamical response of the envelope for higher companion masses. This leads to rapid expansion and therefore the transfer of orbital energy from the companion onto the envelope gas by tidal drag is less efficient. For very low companion masses, however, the dynamical response may eventually become so weak that the envelope expands only little and insufficient recombination energy is released for envelope ejection \citep[see][]{kramer2020a}. 
We therefore anticipate an optimal envelope removal efficiency at intermediate mass ratios. 
Extrapolating from our data, we expect complete envelope ejection in all considered cases. It will be achieved $\SI{7.4}{\year}$ (O.25), $\SI{8.2}{\year}$ (O.50) and $\SI{9.4}{\year}$ (O.75) after the beginning of the simulation. 

\begin{table*}
\centering
\caption[Unbound mass and energies with different companions.]{Fractions of unbound mass, $f_{ej} = M_{ej}/M_e$, and energies for the different companions at the end of the runs. 
The relative energy error of the simulation is estimated by $e_M = \Delta E_{tot}/|E_{pot}^{\mathrm{ini}}|$ with the absolute energy error $\Delta E_{tot}=E_{tot}^{\mathrm{fin}}-E_{tot}^{\mathrm{ini}}$.} 
\begin{tabular}{l c c c c c c c c c }
\toprule
Model & $f_{ej,kin}$ & $f_{ej,therm}$ & $f_{ej,OPAL}$ & $f_{inorb}$ & $e_M$ & $\Delta E_{orb}$ & $\Delta E_{therm}$ & $\Delta E_{recomb}$ & $\Delta E_{tot}$\\
&&&&&&$[\SI{e46}{\erg}]$&$[\SI{e46}{\erg}]$&$[\SI{e46}{\erg}]$&$[\SI{e46}{\erg}]$\\
\midrule
I.25 & $\SI{0.16}{}$ & $\SI{0.17}{}$ & - & $\SI{0.00}{}$ & $\SI{0.03}{}$ & $\SI{-1.21}{}$ & $\SI{-1.25}{}$ & - & $\SI{0.09}{}$\\
I.50 & $\SI{0.20}{}$ & $\SI{0.21}{}$ & - & $\SI{0.01}{}$ & $\SI{0.02}{}$ & $\SI{-1.12}{}$ & $\SI{-1.20}{}$ & - & $\SI{0.08}{}$\\
I.75 & $\SI{0.06}{}$ & $\SI{0.08}{}$ & - & $\SI{0.04}{}$ & $\SI{0.01}{}$ & $\SI{-0.91}{}$ & $\SI{-1.07}{}$ & - & $\SI{0.06}{}$\\

O.25 & $\SI{0.91}{}$ & $\SI{0.98}{}$ & $\SI{1.00}{}$ & $\SI{0.00}{}$ & $\SI{0.07}{}$ & $\SI{-0.95}{}$ & $\SI{-1.22}{}$ & $\SI{-0.98}{}$ & $\SI{0.23}{}$\\
O.50 & $\SI{0.86}{}$ & $\SI{0.94}{}$ & $\SI{0.98}{}$ & $\SI{0.02}{}$ & $\SI{0.05}{}$ & $\SI{-0.86}{}$ & $\SI{-1.16}{}$ & $\SI{-0.95}{}$ & $\SI{0.19}{}$\\
O.75 & $\SI{0.75}{}$ & $\SI{0.87}{}$ & $\SI{0.95}{}$ & $\SI{0.04}{}$ & $\SI{0.09}{}$ & $\SI{-0.62}{}$ & $\SI{-1.05}{}$ & $\SI{-0.88}{}$ & $\SI{0.17}{}$\\ 
\bottomrule
\end{tabular}
\newline
\begin{flushleft}
$f_{ej,kin}$: unbound mass fraction of the envelope with kinetic energy exceeding gravitational binding energy, $e_{grav} + e_{kin} > 0$; \\
$f_{ej,therm}$: unbound mass fraction of the envelope with kinetic plus thermal energy, $e_{grav} + e_{kin} + e_{therm} > 0$; \\
$f_{ej,OPAL}$: unbound mass fraction of the envelope with kinetic plus total internal energy (OPAL \eos only), $e_{grav} + e_{kin} + e_{int} > 0$;\\ 
$f_{inorb}$: mass fraction of the envelope within the orbit of the two cores;\\
$E_{orb} = (M_c + f_{inorb} M_e)M_2/(2a)$: orbital energy, not accounting for the mass outside of the orbit;\\
$E_{therm}$: thermal energy, determined from the \eos;\\
$E_{recomb}$: recombination energy for simulations employing the OPAL \eos only.
\end{flushleft}
\label{table:Lcompeal}
\end{table*}

\subsection{Final-to-initial-separation ratio}
\label{sec:fin-sep}

\begin{figure}
\centering
\includegraphics[width=\columnwidth]{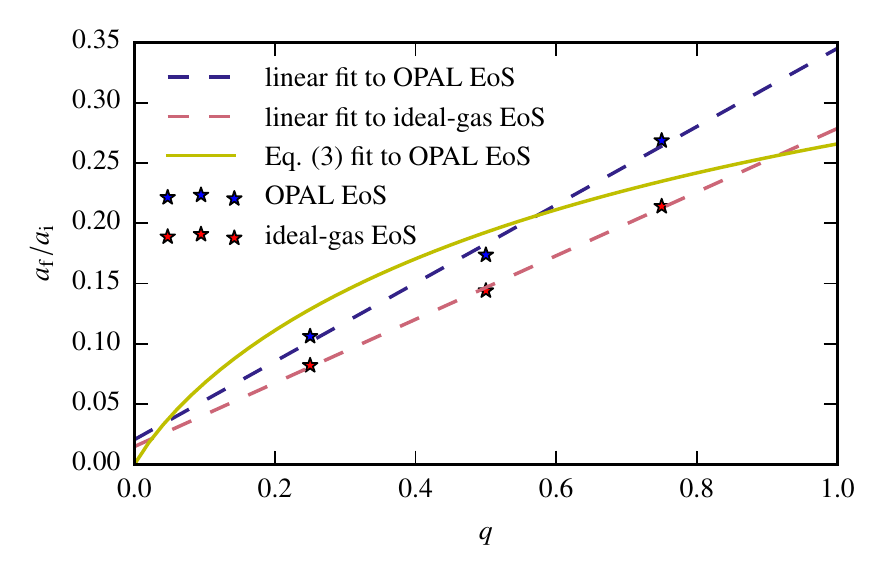}
\caption[Final orbital configuration.]{Final orbital configuration as a function of $q$. Linear fits (dashed lines) to the three data points per \eos (stars) are shown in blue (OPAL \eos) and red (ideal-gas \eos). A fit to Eq.~(\ref{eq:dt2000}) with $\alpha \lambda = 2.1$ is shown in yellow.}
\label{fig:afaiq}
\end{figure}

The ratio of the final to the initial orbital separation given by the formalism in Eq.~(4) of \citet{dewi2000a} is 
\begin{equation}
\label{eq:dt2000}
\frac{a_f}{a_i} = \frac{M_c M_2}{M_c + M_e} \frac{1}{M_2 + 2 M_e / (\alpha \lambda r_L)}.
\end{equation}
A fit of this relation to the data of our OPAL \eos-based simulations with $\alpha \lambda = 2.1$ is shown in Fig.~\ref{fig:afaiq}. Surprisingly, however, a linear relation seems to fit the simulation data better.
We perform a linear regression on the ratio of the orbital separations depending on $q$, the initial mass ratio at the onset of the CE, and find for the OPAL-\eos runs
\begin{equation}
\frac{a_f}{a_i} = (0.32 \pm 0.03)\, q + (0.02 \pm 0.02), 
\end{equation}
whereas we obtain 
\begin{equation}
\frac{a_f}{a_i} = (0.26 \pm 0.01)\, q + (0.01 \pm 0.01) 
\end{equation}
for the ideal-gas \eos runs. A comparison run with $q=0.5$ and a larger initial separation ($p=0.8$) was performed, leading to a ratio of $a_f/a_i = 48\,R_{\odot}/314\,R_{\odot} = 0.15$. This is in the range of our fit and confirms that it works well also for different initial separations.

\subsection{$\alpha$-formalism}
\label{sec:alpha-res}

For the OPAL-\eos runs, we anticipate that the entire envelope is ejected such that we can employ the $\alpha$-formalism of \citet{webbink1984a}. 
In population synthesis models, the envelope-ejection efficiency $\alpha_{CE}$ of CE phases is usually parametrized in terms of the released orbital energy $\Delta E_{orb}$ and the binding energy of the envelope of the primary star $E_{bin}$,
\begin{equation}\label{eq:alpha}
\alpha_{CE} = \frac{E_{bin}}{\Delta E_{orb}}.
\end{equation}
This parametrization serves for determining the final orbital separations of compact binary systems formed in CEE. The binding energy $E_{bin}$ is usually described using the structural parameter $\lambda$ (see Sect.~\ref{sec:relaxresults}). 
The released orbital energy is approximately given by
\begin{equation}
\Delta E_{orb} \approx - G \left[ \frac{M_c M_2}{2 a_f} - \frac{(M_c + M_e)M_2}{2 a_i} \right],
\end{equation}
and thus
\begin{equation}
\label{eq:alphalong}
- G \frac{M_e \left( \frac{1}{2}M_e + M_c \right) }{\lambda R} \approx - \alpha_{CE} G \left[ \frac{M_c M_2}{2 a_f} - \frac{(M_c + M_e)M_2}{2 a_i} \right].
\end{equation}
With an ad-hoc choice of $\alpha_{CE}$ a value for $a_f$ can be determined. 

Our simulations allow to directly determine a value $\alpha= E_{bin}/\Delta E_{orb}$. We choose the definition of $\alpha$ as in Eq.~(\ref{eq:alpha}) so that we are as close as possible to population-synthesis models. Therefore, $\alpha$ primarily serves to determine the final orbital separations and has -- in this form -- little significance for the envelope ejection. 
The resulting values for $\alpha$ are calculated either using the potential energy only or using the potential and the entire internal energy (\ie, $E_g$ and $E_b$ from Table~\ref{table:lambda}). They are given in Table~\ref{table:aLcompeal}. When only accounting for the gravitational energy, the ejection seems to need more energy than taken from the orbit ($\alpha_g > 1$), but when taking into account also the internal energy, the efficiency $\alpha_b$ is less than one. 
Since we assume complete envelope ejection, computing $\alpha$ is meaningful only for the OPAL-\eos models. 

\begin{table}
\centering
\caption[Values of $\alpha$ with different companions.]{Values of $\alpha$ for the OPAL \eos with different companions, calculating $\alpha_g$ with $E_{bin}=E_g$ (excluding the internal energy) and $\alpha_b$ with $E_{bin}=E_b$ (including the internal energy) from Table~\ref{table:lambda}.}
\begin{tabular}{l c c}
\toprule
model & $\alpha_g$ & $\alpha_b$\\
\midrule
O.25 & 3.00 & 0.34\\
O.50 & 3.31 & 0.37\\
O.75 & 4.57 & 0.51\\
\bottomrule
\end{tabular}
\label{table:aLcompeal}
\end{table}

\section{Discussion}
\label{sec:discussion}

Previous high-resolution three-dimensional simulations with \arepo by \citet{ohlmann2016a}, conducted with a $2\,M_{\odot}$ RG star as primary, observe shear instabilities between the individual layers of the spiral structure characteristic for CE interaction. In their simulations, the instability-driven flow patterns grow in size and eventually large-scale flow instabilities wash out the spiral structure completely. Our simulations including the release of recombination energy, however, show no large-scale flow instabilities, only small-scale instabilities arise within the spiral arms. The spiral structure is preserved until large parts of the envelope are ejected. We interpret this effect as a suppression of the instabilities due to a stronger expansion of the envelope gas. 

In our simulations without recombination energy, only up to about $\SI{20}{\percent}$ of the envelope material is expelled. A comparison with the mere $8\%$ mass loss in the simulation of \citet{ohlmann2016a} with a RG primary star (but an otherwise identical numerical approach) confirms the expected effect of a more efficient envelope ejection for a less tightly bound AGB primary. Differences in the setup of the models, the numerical methods, and the achieved numerical resolutions render a quantitative comparison with other studies difficult. Nonetheless, the $23$--$\SI{31}{\percent}$ envelope ejection found by \citet{sandquist1998a} in simulations with more massive AGB primaries of $3\,M_{\odot}$ and $5\,M_{\odot}$ and the $\SI{25}{\percent}$ of unbound envelope mass with an AGB primary of $3.05\,M_{\odot}$ ($3.5\,M_{\odot}$ ZAMS mass) reported by \citet{staff2016b} fall into the same ballpark.

\citet{chamandy2020a} follow CEE in a simulation with an AGB primary star for 20~orbital revolutions and find an envelope ejection of about $\SI{11}{\percent}$. By the end of their simulation, however, the unbound mass is still increasing. They extrapolate that a complete envelope ejection is possible in less than ten years, provided the mass loss rate does not change significantly. Our simulation I.50 (which, like that of \citealp{chamandy2020a}, ignores recombination energy release) follows the evolution up to 70 orbits. By then, the mass loss rate has decreased substantially. The unbound mass still grows slightly, but with a decreasing rate. It seems questionable if an ejection of more than $\sim \SI{25}{\percent}$ of the envelope mass is possible. The reason for the strong decrease in mass loss after the plunge-in phase is an expansion of envelope gas so that little material is left inside the orbit of the stellar cores. This makes the transfer of orbital energy to the gas inefficient. We cannot, however, exclude the establishment of a self-regulated inspiral phase \citep{meyer1979a,podsiadlowski2001a}, where the envelope contracts on a thermal time scale and episodes of more efficient energy transfer followed by another phase of expansion and contraction lead to a slow loss of the envelope.

As for the case of CEE with RG primary stars \citep{nandez2015a, prust2019a, reichardt2020a}, the question of accounting for the release of recombination energy turns out to be decisive for the success of envelope ejection also with AGB primaries. Our simulations strongly indicate that a complete envelope removal is possible provided this energy is indeed transferred to the envelope material. Contrary to \citet{grichener2018a}, \citet{ivanova2018a} argues that only a negligible fraction of recombination energy is radiated away and most of it can be used to eject the envelope. This is supported by our simulations, where we find that during the evolution most of the remaining ionization energy is located in regions of high optical depths.

\cite{fragos2019a} simulated the complete evolution of a binary system with a $12\,M_{\odot}$ red supergiant primary star using a 1D hydrodynamic \mesa model. Including recombination energy, they conclude that most of the envelope is unbound. However, as in our case, the thermal energy contribution is more important than recombination.
\cite{ricker2019b} studied CEE with $82.1\,M_\odot$ red supergiant stars with and without radiative transfer in 3D for the first time and found that for these specific objects recombination energy does not help in envelope removal. This is no contradiction to our result as the structures of the primary stars in the considered mass ranges are quite different \citep[see, \eg, Figure~5 in][]{kruckow2016a}.

Outflow and unbinding of envelope gas driven by recombination continue after the plunge-in phase in our case. Other mechanisms, such as pre-plunge-in ejection and ejection triggered by a contraction of the circum-binary envelope \citep{ivanova2016a}, accretion onto the companion, and dust formation are not necessarily required for successful envelope ejection, but may still play a role. The mass loss rate at the end of our simulations with the OPAL \eos is stable over many orbits. A simple extrapolation yields complete envelope removal within about ten years. The similarity to the value estimated by \citet{chamandy2020a} is probably sheer coincidence, because the physical effects responsible for envelope ejection are different in the models: \citet{chamandy2020a} do not account for recombination effects and extrapolate the mass unbinding found in the initial plunge-in phase. 

In our simulations, the core system approaches a final and steady orbital separation. The orbital evolution is similar in the runs accounting for the release of ionization energy and those ignoring this effect. It is largely determined by the initial plunge-in phase. Later recombination processes occur primarily in the outer parts of the envelope and are important for mass ejection, but have less (although not negligible) effect on the inner core binary. Being a key quantity for the future evolution of the eventually formed close compact binary system, we attempt to determine how the final core separation depends on parameters of the initial setup. Remarkably, our simulations suggest a linear relation between the final-to-initial-separation ratio and the mass ratio in the original binary. However, there is no such simple linear relation between the eccentricity and the mass ratio. We provide values for $\alpha$ from our simulations employing the OPAL \eos. We find agreement when comparing our values from the OPAL simulations resulting in $\lambda \alpha$ values of 0.96, 1.06, and 1.46 to the range of 0.75 to 1.27 reported by \cite{nandez2016a}. 
As discussed by \citet{iaconi2019a}, our values for the final separations and $\alpha$ are upper limits. In contrast to the hypothesis of \citet{iaconi2019a} and \citet{reichardt2020a}, however, the final separations determined in our simulation are systematically larger when ionization effects are accounted for. This is explained by the fact that recombination energy is already released during the inspiral and the resulting envelope expansion stalls the orbital shrinking earlier. 

The mass ratios $q$ considered in our study have been chosen without referring to a particular model of CE initialization thus raising the question of whether the corresponding systems would enter CEE in the first place. 
\cite{ge2010a,ge2015a,ge2020a} use an adiabatic mass-loss model to systematically determine critical $q$ values for mass transfer to proceed on a dynamical timescale. They assume conservation of mass and orbital angular momentum, and neglect stellar winds. 
\cite{ge2020b} conclude that $q = 3/4$ may be close to the threshold beyond which stable mass transfer is expected instead of a CE episode for a star like ours. This indicates that our study covers the relevant range in $q$.
The outcome of their CEE with a $3.2\,M_{\odot}$ AGB primary star (Figure~7 of \citealp{ge2010a}) appears similar to a binary system with a $12\,M_{\odot}$ red supergiant primary star as shown in Figure~4 of \cite{fragos2019a}.
 
The final orbital separations determined from observations of post-CE 
binaries show a significant scatter, and deriving the initial system 
parameters is challenging and introduces uncertainties. Although our 
simulation results are close to some observed systems
\citep[see][]{iaconi2019a}, the general tendency in those seems to point
toward smaller $a_f$ than obtained here. This certainly 
warrants further study. In particular, higher-mass AGB primary stars
are expected to lead to a deeper inspiral of the companion because more
envelope material has to be ejected. 

The final core separations are smaller for lower mass ratios in the AGB systems studied by \cite{sandquist1998a}. Our results confirm this trend.
Starting at a core distance of $124\,R_{\odot}$ in a system with mass ratio $q=0.55$, \cite{chamandy2020a} find a final separation of $16\,R_{\odot}$. The resulting $a_f/a_i=0.13$ is close to our fit for the ideal-gas runs, even though further spiral-in is probable. 
\citet{staff2016b} find no convergence with resolution when comparing their runs ``4'' and ``4hr'', but their separation ratio falls below our fit. 

Our results raise the question of the final fate of the simulated systems. The unbound mass fraction is larger for smaller $q$ by the end of the respective simulations, but it is still increasing in all OPAL runs. We cannot follow the evolution for longer with confidence with our current numerical methods, because the energy-error rate exceeds the recombination-energy-release rate in the system and we can no longer decide whether a further envelope unbinding is physical or caused by numerical errors. Moreover, additional physical effects may become important in the late phases of the evolution modeled here. With further envelope expansion, recombination may not proceed in optically thick regions and the energy may be radiated away instead of aiding mass ejection. A counteracting effect would be the above-mentioned dust formation that may revive energy transfer to envelope material. If the envelope ejection remains incomplete, recurrent CE episodes seem possible until full removal of the material is achieved. In this case, the orbital separation may again shrink (although probably not by much since the binding energy of the remaining material is low) and this would affect its relation to the initial mass ratio and the $\alpha$ values determined here.

\section{Conclusions}
\label{sec:conclusions}

We present simulations of common envelope interactions of a low-mass
AGB primary star ($1.2\,M_{\odot}$ ZAMS mass; $173\,R_{\odot}$ radius
at onset of common-envelope evolution) with different companions using
the \arepo code. Our simulations follow the common-envelope evolution up to about $100$ orbits of the core binary.

Despite the lower binding energy of the AGB star compared to a RG, envelope ejection stalls below 20\% of its mass when not accounting for ionization energy release. By employing the OPAL \eos, we include this effect in our simulations, effectively assuming that the recombination energy liberated in the expanding envelope is thermalized locally. In this case, our simulations indicate that complete envelope ejection is possible. 

Comparing simulations with different companion masses, we find that less massive companions spiral in deeper into the envelope. By the time our simulations terminate, they have led to a more complete envelope ejection in our simulations accounting for ionization effects. However, because the unbound mass fraction keeps increasing almost linearly in our simulations including ionization effects, the final envelope ejection is expected to be complete in all cases under the assumption of local thermalization of released recombination energy. Because of the stronger dynamical response of the envelope for more massive companions and the lack of expansion and recombination energy release for too low-mass companions, we expect an optimal envelope removal efficiency at intermediate mass ratios of the two stars. 

At least in the important early phases recombination energy is released in optically thick regions, such that it cannot be radiated away and is bound to support mass loss. This supports the assumptions made in our simulations. For a final verdict on envelope ejection, however, a proper treatment of the release of energy from recombination processes and the associated radiation requires the inclusion of radiative transfer.

Our simulations indicate that a simple linear relation may provide a satisfactory fit to the dependence of the final orbital separation of the core binary on the initial mass ratio between secondary and primary star. This has to be confirmed with further simulations exploring a wider parameter space. 

\begin{acknowledgements}
The work of CS and FKR was supported by the Klaus Tschira Foundation.
Parts of this work were performed on the computational resource ForHLR~I
funded by the Ministry of Science, Research and the Arts Baden-W{\"u}rttemberg and
DFG (``Deutsche Forschungsgemeinschaft'').
For data processing and plotting, NumPy \citep{oliphant2006a} and SciPy \citep{virtanen2020a}, IPython \citep{perez2007a}, and Matplotlib \citep{hunter2007a} were used. 
We thank the anonymous referee, Noam Soker and Matthias Kruckow for helpful comments.
\end{acknowledgements}

\bibliographystyle{aa} 

\end{document}